# Accounting for the Uncertainty in the Evaluation of Percentile Ranks

*Sir,*

In a recent paper entitled "Inconsistencies of Recently Proposed Citation Impact Indicators and how to Avoid Them," Schreiber (2012) proposed (*i*) a method to assess tied ranks consistently and (*ii*) fractional attribution to percentile ranks in the case of relatively small samples (e.g., for *n* < 100). For example, a journal may contain a limited number of (e.g., ten) reviews in a year. In that case, the highest possible percentile in this category in this year would be based on (9/10) or 90%. Rousseau (2012) proposes to define this highest possible rank as 100% by including the ranked paper in the ranking, and thus to consider (10/10) as the highest possible rank. Leydesdorff & Bornmann (in press) argue that this definition of a percentile is not compatible with standard definitions (Hyndman & Fan, 1996) and too easily leads to an upward drift in percentile rankings so that non-cited papers could be greatly overestimated in a citation analysis (Zhou *et al.*, in preparation).

Since percentile rankings are frequently used in evaluation studies, for example, as excellence indicators in university rankings—both the Leiden rankings and the Scimago institute rankings use the top-10% most-highly-cited papers as basis for the ranking (Bornmann *et al.*, 2012)—a further proliferation of possible definitions of "percentile ranks" would be confusing. From this perspective, Schreiber's contribution is timely: his solution to the problem of how to handle tied ranks is convincing, in my opinion (cf. Pudovkin & Garfield, 2009). The fractional attribution, however, is computationally intensive and cannot be done manually for even moderately large batches of documents. I therefore decided to implement both proposals as an additional option into the routine isi2i3.exe (available at http://www.leydesdorff.net/software/i3) which enables the user to convert "times cited" values of documents in batch mode into percentile ranks.

In addition to using hundred percentile ranks, the *Science and Engineering Indicators* of the U.S. National Science Board (e.g., 2012) has for a decade used six classes in the evaluation: top-1%, top-5%, top-10%, top-25%, top-50%, and bottom-50% (Bornmann & Mutz, 2011). Leydesdorff *et al*. (2011) suggested distinguishing the use of this or any other normative evaluation scheme (e.g., Adams *et al*., 2000) from the analytical question of how to attribute percentile-values. The classifications—for example, in these six percentile ranks classes (*PR6*)—can be considered as (possibly different) aggregation rules. The aggregation may introduce a non-linearity, as in the case of *PR6*.

In my opinion, Schreiber (2012) confused the normative and the analytical question. He attributed scores fractionally to *PR6*, and thus missed the point that fractional attribution at the level of hundred percentiles (*PR100*)—or equivalently quantiles as the continuous random variable—is only a linear, and therefore much less complex problem. In the case of a document set of $n = 8$, for example, the highest possible quantile is $100 * (7/8) = 87.5\%$; or 100% using Rousseau's counting rule. The interval 87.5-100 can thus be considered as the uncertainty. Schreiber proposed to attribute this interval proportionally to *PR6* as follows: the interval 87.5-90% to the top-25% (contributing to the score fractionally with 2.5/12.5), the interval between 90-95% to the top-10% (5/12.5), the interval 95-99% to the top-50% (4/12.5), and 99-100% to the top-1% (1/12.5). This leads to a sum score as follows: $(2.5/12.5) * 3 + (5/12.5) * 4 + (4/12.5)$

* 5 + (1/12.5) * 6 = 53.5/12.5 = 4.28. After rounding, this document can thus be considered as belonging to the class between 90 and 95% with *PR6* = 4.

The more general solution of *PR 100* (or equivalently, quantiles) provides us with the same score. The interval 87.5-100 is then linear—each percentage point contributes equally—and the score is (87.5 + 100)/2 = 93.75, and thus *PR6* = 4. (However, the correspondence in outcome between the two procedures is incidental and not necessary because a non-linearity is involved in the binning.) The specification of the uncertainty between 87.5 and 100 clarifies that the best attribution given the information available would be 93.75%. The uncertainty is equal to $1/N$ in which $N$ is the number of documents. Using the counting rule of Leydesdorff & Bornmann (in press) one would add half of the uncertainty (6.25 percent points) to the result (87.5 + 6.25 = 93.75) and using the counting rule of Rousseau (2012) one would subtract: 100 − 6.25 = 93.75. It thus follows from either procedure that *PR6* = 4 and *PR 100* = 93.75. Other evaluation schemes—for example, in terms of quartiles—can be translated into respective aggregation rules of the quantile values (Adams *et al*., 2000). The results can be tested for significance against the expected value of 50 as the median for percentiles or quartiles (e.g., using Wilcoxon signed-rank test in SPSS; Leydesdorff *et al*., 2011), and against the value of 1.91 for the random attribution to *PR 6* (Bornmann & Mutz, 2011).

Given the generality of this solution, I added it as the default option to the routine isi2i3.exe, while leaving the two other options (our previous definition and that of Rousseau) available. Thanks to Schreiber's two proposals, but with this simplification, the problem of specifying quantiles in small groups is now, in my opinion, solved. Different evaluation schemes can be used independently from the analytical processing. Because the uncertainty in the attribution is equal to $1/N$—and the correction terms accordingly $1/2N$—uncertainty vanishes with larger values of $N$.

Loet Leydesdorff [*]

April 6, 2012

**References**
Adams, J., Cooke, N., Law, G., Marshall, S., Mount, D., Smith, D., . . . Stephenson, J. (2000). The Role of Selectivity and the Characteristics of Excellence. Report to the Higher Education Funding Council for England (pp. 76). Leeds, UK / Philadelphia, PA; at http://www.evidence.co.uk/downloads/selectivity-report.pdf: Evidence Ltd. Higher Education Policy Unit, University of Leeds/ ISI.
Bornmann, L., & Mutz, R. (2011). Further steps towards an ideal method of measuring citation performance: The avoidance of citation (ratio) averages in field-normalization. *Journal of Informetrics, 5*(1), 228-230.
Hyndman, R. J., & Fan, Y. (1996). Sample quantiles in statistical packages. *American Statistician*, 361-365.
Leydesdorff, L., & Bornmann, L. (in press). Percentile Ranks and the Integrated Impact Indicator (*I3*). *Journal of the American Society for Information Science and Technology*; preprint available at http://arxiv.org/abs/1112.6281.
Leydesdorff, L., Bornmann, L., Mutz, R., & Opthof, T. (2011). Turning the tables in citation analysis one more time: Principles for comparing sets of documents *Journal of the American Society for Information Science and Technology, 62*(7), 1370-1381.
National Science Board. (2012). *Science and Engineering Indicators*. Washington DC: National Science Foundation; available at http://www.nsf.gov/statistics/seind12/.
Pudovkin, A. I., & Garfield, E. (2009). Percentile Rank and Author Superiority Indexes for Evaluating Individual Journal Articles and the Author's Overall Citation Performance. *CollNet Journal of Scientometrics and Information Management, 3*(2), 3-10.
Rousseau, R. (2012). Basic properties of both percentile rank scores and the I3 indicator. *Journal of the American Society for Information Science and Technology, 63*(2), 416-420.
Schreiber, M. (in press). Inconsistencies of Recently Proposed Citation Impact Indicators and how to Avoid Them. *Journal of the American Society for Information Science and Technology*; preprint available at http://www.arxiv.org/abs/1202.3861.
Zhou, P., Ye, F. Y., & Zhong, Y. (in preparation). The Combined Impact Indicator: Another option for measuring academic impact.

---

[*] University of Amsterdam, Amsterdam School of Communication Research (ASCoR), Kloveniersburgwal 48, 1012 CX Amsterdam, The Netherlands; loet@leydesdorff.net ; http://www.leydesdorff.net